\documentclass[twocolumn]{article}
\usepackage[dvipdfm]{graphicx}

\begin{document}
\bibliographystyle{jplain} 
%


\title{{\bf  Implementing Continuation based language in GCC}}

\author{ Shinji KONO, Kento YOGI \\[5mm]
e-mail:kono@ie.u-ryukyu.ac.jp\\
Information Engineering, University of the Ryukyus\\
Nishihara-cyo 1, Okinawa, 903-01, Japan}
\date{2008 Apr 13}


\twocolumn[   
\maketitle{}


\mbox{}

We have implemented C like Continuation based programming language.
Continuation based C, CbC was implemented using micro-C on various architecture, 
and we have tried several CbC programming experiments. 
Here we report new implementation
of CbC compiler based on GCC 4.2.3. Since it contains full C capability, we
can use CbC and C in a mixture.


]

\section{ A Practical Continuation based Language}

If CPS theory is successful, it should also be working well in practical area.
Our idea is simple. How about a programming language which has continuation
passing style only? How about it runs as fast as current GNU C compiler?

Instead of creating complete new programming language, we designed a lower
language of C, so called Continuation based C, here after CbC. Using CPS
transformation like method, we can compile C into CbC, that is, we have some
kind of backward compatibility.

We have implemented CbC using micro-C on various architecture, and we have
tried several CbC programming experiments. Here we report new partial implementation
of CbC compiler\cite{cbc-sourceforge} based on GCC 4.2.3\cite{gcc}. Since it contains full C capability, we
can use CbC and C in a mixture, so when call the mixture C with C, here after CwC.

First we show CbC language overview.

\section{ Continuation based C}

CbC's basic programming unit is a code segment. It is not a subroutine, but it
looks like a function, because it has input and output. We can use C struct
as input and output interfaces.

{\small
\begin{verbatim}
   struct interface1  { int i; };
   struct interface2  { int o; };

   __code f(struct interface1 a) { 
       struct interface2 b; b.o=a.i;  
       goto g(b); 
   }

\end{verbatim}
}

\begin{figure}[htb]
\begin{center}
\includegraphics[width=6cm]{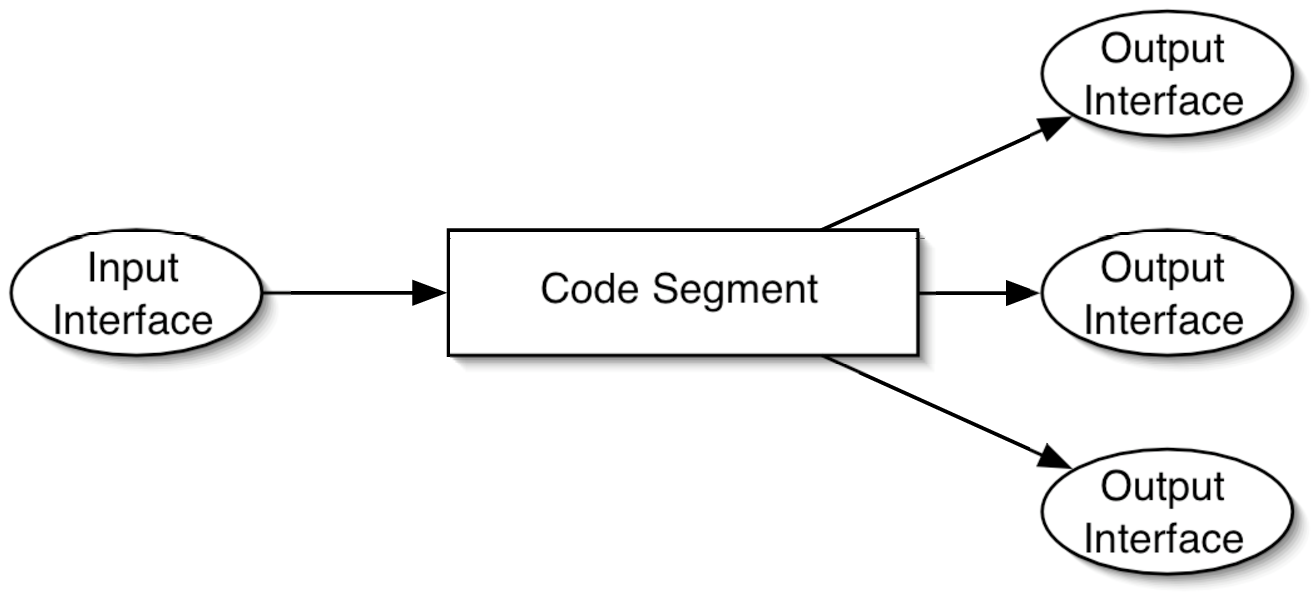}
\caption{code}
\end{center}
\label{code}
\end{figure}

In this example, a code segment
\verb+f+ has \verb+input a+ and sends \verb+output b+ to a code segment \verb+g+.
There is no return from code segment \verb+b+, \verb+b+ should call another
continuation using \verb+goto+. Any control structure in C is allowed in CwC
language, but in case of CbC, we restrict ourselves to use \verb+if+ statement
only, because it is sufficient to implement C to CbC translation. In this case,
code segment has one input interface and several output interfaces (fig.\ref{code}).

\verb+__code+ and parameterized global goto statement is an extension of
Continuation based C. Unlike \verb+C--+ \cite{cminusminus}'s parameterized goto,
we cannot goto into normal C function.

\subsection{ Intermix with C}

In CwC, we can go to a code segment from a C function and we can call C functions
in a code segment. So we don't have to shift completely from C to CbC. The later
one is straight forward, but the former one needs further extensions.

{\small
\begin{verbatim}
   void *env;
   __code (*exit)(int);

   __code h(char *s) { 
        printf(s);
        goto (*exit)(0),env;
   }

   int main() {
        env  = __environment;
        exit = __return;
        goto h("hello World\n");
   }

\end{verbatim}
}

In this hello world example, the environment of \verb+main()+
and its continuation is kept in global variables. The environment
and the continuation can be get using \verb+__environment+,
and \verb+__return+. Arbitrary mixture of code segments and functions
are allowed (in CwC). The continuation of \verb+goto+ statement 
never returns to original function, but it goes to caller of original
function. In this case, it returns result 0 to the operating system.

\section{ What's good?}

CbC is a kind of high level assembler language. It can do several
original C language cannot do. For examples,

{\small
\begin{verbatim}
    Thread Scheduler 
    Context Switch
    Synchronization Primitives
    I/O wait semantics

\end{verbatim}
}

are impossible to write in C. Usually it requires some help of
assembler language such as \verb+__asm+ statement extension which is
of course not portable.

\subsection{ Scheduler example}

We can easily write these things in CbC, because
CbC has no hidden information behind the stack frame of C. 
A thread simply go to the scheduler,

{\small
\begin{verbatim}
    goto scheduler(self, task_list);

\end{verbatim}
}

and the scheduler simply pass the control to the next
thread in the task queue.

{\small
\begin{verbatim}
    code scheduler(Thread self,TaskPtr list)
    {
        TaskPtr t = list;
        TaskPtr e;
        list = list->next;
        goto list->thread->next(list->thread,list);
    }

\end{verbatim}
}

Of course it is a simulator, but it is an implementation
also. If we have a CPU resource API, we can write real multi
CPU scheduler in CbC.

This is impossible in C, because we cannot access the hidden
stack which is necessary to switch in the scheduler. In CbC,
everything is visible, so we can switch threads very easily.

This means we can use CbC as an executable  specification 
language of OS API.

\subsection{ Self Verification}

Since we can write a scheduler in CbC, we can also enumerate
all possible interleaving of a concurrent program. We have
implement a model checker in CwC. CbC can be a self verifiable
language\cite{kono08a}.

SPIN\cite{holzmann97model} is a very reliable model checker, but it have to
use special specification language PROMELA. We cannot directly
use PROMELA as an implementation language, and it is slightly
difficult to study its concurrent execution semantics including
communication ports.

There are another kind of model checker for real programming
language, such as Java PathFinder\cite{havelund98model}. Java PathFinder use
Java Virtual Machine (JVM) for state space enumeration which
is very expensive some time.

In CbC, state enumerator itself is written in CbC, and its concurrency
semantics is written in CbC itself. Besides it is very close
to the implementation. Actually we can use CbC as an implementation
language. Since enumerator is written in the application itself, we
can perform abstraction or approximation in the application specific
way, which is a little difficult in Java PathFinder. It is possible
to handle JVM API for the purpose, although.

We can use CPS transformed CbC source code for verification, but
we don't have to transform all of the source code, because CwC
supports all C constructs. (But not in C++... Theoretically it is
possible with using cfront converter, it should be difficult).

\subsection{ As a target language}

Now we have GCC implementation of CbC, it runs very fast. Many
popular languages are implemented on top of C. Some of them
uses very large switch statement for the byte code interpreter.
We don't have to use these hacks, when we use CbC as an implementation
language. 

CbC is naturally similar to the state charts. It means it is very
close to UML diagrams. Although CbC does not have Object Oriented
feature such as message passing nor inheritance, which is not
crucial in UML.

\section{ Transformation (C2CbC)}

Conversion from C to CbC is straight forward, but it generates
a lot of code segments. Since CbC does not have heap management
itself, the stack area have to be allocated explicitly.

We find GCC can perform better optimization in translated code
segment. We will discuss it later.

We have an easy implementation of C to CbC compilation, but it is
not a practical level, but we need good converter for backward
compatibility.

We can also consider possible conversion from C++ to CbC.
In this case, all hidden operation in C++ should become explicit,
for examples, object allocations and deallocations in the stack,
handling of auto pointer and so on.

\section{ GNU CC implementation}

So how to implement CwC in GCC.
The idea itself is simple\cite{kono02f}, forcing C tail call elimination for all
code segment. 

But before GCC version 4.x, tail call elimination (here after TCE) is not so cleanly
implemented , it is very difficult to implement it. But in GCC 4.x,
basically TCE can be applied for all possible functions.

\verb+__code+ is implemented as a new type keyword in GCC. You may think
\verb+__code+ is an attribute of a function, which means that the function
can call in tail call elimination only.

Because of this implementation, we can actually call code segment as a
normal function call.

\subsection{ How to force tail call elimination}

There many enable conditions for tail call elimination, for example,
there should be no statement after tail call, return value type have to
be the same, arguments size should be compatible, and so on. We find
almost half of lines in \verb+calls.c+ spends to check TCE possibilities.

Our conclusion is this. It is not practical to make sure to pass all the
TCE tests, instead, we write TCE only version of \verb+expand_call()+
separately in 783 lines. 

{\small
\begin{verbatim}
    4463   18527  145469 calls.c          
        expand_call() for function
     783    2935   23651 cbc-goto.h       
        expand_cbc_goto() for code segment

\end{verbatim}
}

All code segment has the same virtual argument size and void return type,
that is argument register or argument value in the memory is shared among
all code segments. This leads a problem.

\subsection{ Parallel Assignment}

Consider the next code,

{\small
\begin{verbatim}
    __code carg4(struct arg args0,struct arg args1,
         int i, int j,int k,int l)
    {
        goto carg5(args1,args0,j,k,l,i);
    }

\end{verbatim}
}

In this case, simple sequential assignments does not work. It
override \verb+args1+ or \verb+args0+. In normal function case,
GCC simply give up TCE, and pushes all arguments in new register or
stack area. We are not allowed that. That is we have to implement
parallel assignment in the code segment goto.

This is done by simple copy overlapped arguments in a stack. We
hope to eliminate unnecessary copy during GCC optimization. 

\subsection{ Not yet done}

Currently we have not yet implemented goto with environment and
\verb+__return+, \verb+__environment+.

In some GCC 4.x supported architecture, TCE itself is not
supported in special case. Our method does not work for the
architecture.

Since we made modifications on GCC compiler itself, our method
is GCC version sensitive. We have to do necessary modifications for
coming new version of GCC.

\section{ Result}

Here is our bench mark program.
{\small
{\small
\begin{verbatim}
    f0(int i) {
        int k,j;
        k = 3+i;
        j = g0(i+3);
        return k+4+j;
    }

    g0(int i) {
        return h0(i+4)+i;
    }

    h0(int i) {
        return i+4;
    }
\end{verbatim}
}

}

It is written in C, we perform CPS transformation in several
steps by hands. There are several optimization is possible.

{\small
{\small
\begin{verbatim}
    /* straight conversion case (1) */

    typedef char *stack;

    struct cont_interface { 
      // General Return Continuation
        __code (*ret)();
    };

    __code f(int i,stack sp) {
        int k,j;
        k = 3+i;
        goto f_g0(i,k,sp);
    }

    struct f_g0_interface {  
       // Specialized Return Continuation
        __code (*ret)();
        int i_,k_,j_;
    };

    __code f_g1(int j,stack sp);

    __code f_g0(int i,int k,stack sp) { // Caller
        struct f_g0_interface *c = 
            (struct f_g0_interface *)(
         sp -= sizeof(struct f_g0_interface));

        c->ret = f_g1;
        c->k_ = k;
        c->i_ = i;

        goto g(i+3,sp);
    }

    __code f_g1(int j,stack sp) {  // Continuation 
        struct f_g0_interface *c = 
          (struct f_g0_interface *)sp;
        int k = c->k_;
        sp+=sizeof(struct f_g0_interface);
        c = (struct f_g0_interface *)sp;
        goto (c->ret)(k+4+j,sp);
    }

    __code g_h1(int j,stack sp);

    __code g(int i,stack sp) { // Caller
        struct f_g0_interface *c = 
            (struct f_g0_interface *)(
           sp -= sizeof(struct f_g0_interface));

        c->ret = g_h1;
        c->i_ = i;

        goto h(i+3,sp);
    }

    __code g_h1(int j,stack sp) {  
        // Continuation 
        struct f_g0_interface *c = 
          (struct f_g0_interface *)sp;
        int i = c->i_;
        sp+=sizeof(struct f_g0_interface);
        c = (struct f_g0_interface *)sp;
        goto (c->ret)(j+i,sp);
    }

    __code h(int i,stack sp) {
        struct f_g0_interface *c = 
          (struct f_g0_interface *)sp;
        goto (c->ret)(i+4,sp);
    }

    struct main_continuation { 
        // General Return Continuation
        __code (*ret)();
        __code (*main_ret)();
        void *env;
    };

    __code main_return(int i,stack sp) {
        if (loop-->0)
            goto f(233,sp);
        printf("#0103:%d\n",i);
        goto (( (struct main_continuation *)sp)->main_ret)(0),
               ((struct main_continuation *)sp)->env;
    }
\end{verbatim}
}

}

This is awfully long, but it is straight forward. Several forward
prototyping is necessary, and we find strict prototyping is
painful in CbC, because we have to use many code segments to
perform simple thing.  CbC is not a language for human, but for
automatic generation, verification or IDE directed programming.

We can shorten the result in this way.
{\small
{\small
\begin{verbatim}
    /* little optimized case (3) */

    __code f2_1(int i,char *sp) {
        int k,j;
        k = 3+i;
        goto g2_1(k,i+3,sp);
    }

    __code g2_1(int k,int i,char *sp) {
        goto h2_11(k,i+4,sp);
    }

    __code f2_0_1(int k,int j,char *sp);
    __code h2_1_1(int i,int k,int j,char *sp) {
        goto f2_0_1(k,i+j,sp);
    }

    __code h2_11(int i,int k,char *sp) {
        goto h2_1_1(i,k,i+4,sp);
    }

    __code f2_0_1(int k,int j,char *sp) {
        goto (( (struct cont_interface *)
           sp)->ret)(k+4+j,sp);
    }

    __code main_return2_1(int i,stack sp) {
        if (loop-->0)
            goto f2_1(233,sp);
        printf("#0165:%d\n",i);
        goto (( (struct main_continuation *)sp)->main_ret)(0),
               ((struct main_continuation *)sp)->env;
    }
\end{verbatim}
}

}

In this example, CPS transformed source is faster than
original function call form. There are not so much area for the
optimization in function call form, because function call API
have to be strict.
CbC does not need standard call API other than interface which
is simply a struct and there are no need for register save. (This
bench mark is designed to require the register save).

Here is the result in IA32 architecture (Table.\ref{tab:mc,gcc,compare}). 
Micro-C is our previous implementation in tiny C. \verb+conv1 1+
is function call. \verb+conv1 2+, \verb+conv1 3+ is optimized CPS transformed
source.

\begin{table}[htpb]
\centering
\small
\begin{tabular}{|l|r|r|r|} \hline
& ./conv1 1 & ./conv1 2 &  ./conv1 3 \\ \hline
Micro-C         & 8.97 & 2.19 & 2.73 \\ \hline \hline
GCC             & 4.87 & 3.08 & 3.65 \\ \hline
GCC (+omit)     & 4.20 & 2.25 & 2.76 \\ \hline
GCC (+fast) & 3.44 & 1.76 & 2.34 \\ \hline 
\end{tabular}
\caption{Micro-C, GCC bench mark (in sec)}
\label{tab:mc,gcc,compare}
\end{table}

There are two optimization flag for GCC. \verb+-fomit-frame-pointer+
eliminates frame pointer (\%ebp). The frame pointer itself is
useful in code segment, but it generates unnecessary push and pop or
\verb+leave+ instruction. Using \verb+fastcall+ option, GCC ignore
the standard call convention such as all argument have be on stack
in IA32. In Micro-C implementation, these optimization is naturally
implemented in code segment, so it is faster than GCC without
these options.

But with these options, GCC is faster than Micro-C. Of course,
in more complex source, GCC's complex optimization should work well.

\section{ Conclusion}

We have designed and implemented Continuation based language for practical
use. We have partial implementation of CwC using GCC 4.2.3. Using suitable
optimized options CPS transformed source sometimes runs faster than
original function call version.

This gcc implementation should be portable on all architectures supporting
tail call elimination, but we have tested only on i386 now. 
\bibliography{ref}

\begin{thebibliography}{1}

\bibitem{gcc}
{Free Software Foundation, Inc.}
\newblock {GCC, the GNU Compiler Collection}, March 2008.

\bibitem{havelund98model}
K.~Havelund and T.~Pressburger.
\newblock Model checking java programs using java pathfinder, 1998.

\bibitem{holzmann97model}
Gerard~J. Holzmann.
\newblock The model checker {SPIN}.
\newblock {\em Software Engineering}, Vol.~23, No.~5, pp. 279--295, 1997.

\bibitem{cminusminus}
Norman Ramsey and Simon~Peyton Jones.
\newblock A single intermediate language that supports multiple implementations
  of exceptions.
\newblock In {\em ACM SIGPLAN 2000 Conference on Programming Language Design
  and Implementation}, June 2000.

\bibitem{cbc-sourceforge}
{Shinji KONO}.
\newblock {CbC}, March 2008.

\bibitem{kono02f}
{河野 真治}.
\newblock {継続を基本とした言語CbCのgcc上の実装}.
\newblock  日本ソフトウェア科学会第19回大会論文集, Sep 2002.

\bibitem{kono08a}
{河野　真治 }.
\newblock {検証を自身で表現できるハードウェア、ソフトウェア記述言語
  Continuation based C と>、そのCell への応用}.
\newblock  電子通信学会VLD研究会, March 2008.

\end{thebibliography}

\end{document}